\documentstyle[preprint,prb,aps,amstex,amssymb]{revtex}
\begin{document}

\draft

\title
{\bf Inhomogeneous D-Wave Superconductivity and Antiferromagnetism in a
Two-Dimensional Extended Hubbard Model with Nearest-Neighbor
Attractive Interaction 
}

\author{
 W. P. Su }

\vspace{0.15in}

\address{
Department of Physics and Texas Center for Superconductivity, 
University of Houston, Houston, Texas 77204, USA }
\maketitle

\begin{abstract}

\vspace{0.15in}

{\it To understand the interplay of $d$-wave superconductivity and 
antiferromagnetism, we consider a two-dimensional
extended Hubbard model with nearest neighbor attractive interaction.
The Hamiltonian is solved in the mean-field approximation on a finite
lattice. In the 
impurity-free case, the minimum energy solutions show phase separation
as predicted previously based on free energy argument. The phase separation
tendency implies that the system can be easily rendered inhomogeneous
by a small external perturbation. Explicit solutions of a model including
weak impurity potentials are indeed inhomogeneous in the spin-density-wave
and $d$-wave pairing order parameters. Relevance of the results to the
inhomogeneous cuprate superconductors is discussed.
 }
\end{abstract}
\pacs{PACS  numbers: 74.20.-z, 74.72.-h, 74.25.Ha}

\vspace{0.15in}

\noindent {\bf I. Introduction}\newline

Motivated by the inhomogeneous $d$-wave superconductivity~\cite{pan,how,kan,el}
and antiferromagnetism~\cite{sin,nie} observed in the
cuprates, we have studied a two-dimensional extended Hubbard model with
nearest-neighbor attractive interaction in a previous paper~\cite{wppp}
(hereafter referred to as I). Free energy of the homogeneous
coexisting antiferromagnetic and superconducting state calculated as a
function of band filling shows a region of negative curvature characteristic
of a phase separation system~\cite{put}. In the two-phase region, the zero 
compressibility means that   large fluctuations in density can be easily
induced by small perturbations. Thus the potentials due to the randomly
distributed dopants (impurities) could lead to inhomogeneous 
d-wave superconductivity and antiferromagnetism. In this paper, we solve
the model Hamiltonian on a finite lattice in the mean-field approximation.
Inhomogeneous solutions are indeed obtained as predicted in I.

This paper is organized as follows. First the model Hamiltonian (the tUVW
model) is defined and the mean-field methodology is described in Section II.
Solutions in the dopant free case are examined in Section III.
Inhomogeneous solutions induced by random dopant potentials are presented
in Section IV. Implications of the solutions are discussed in Section V.

\vspace{0.15in}

\noindent {\bf II. The tUVW Model}

The tUVW model~\cite{mic1} is defined as \vspace{0.1in}

\begin{eqnarray}
H=-t\sum_{<ij>\sigma}[c_{i\sigma}^{\dagger}c_{j\sigma}+H.c.]+
U\sum_{i}(n_{i\uparrow}-{\frac{1}{2}})(n_{i\downarrow}-{\frac{1}{2}})+
V\sum_{<ij>}(n_{i}-1)(n_{j}-1)  \nonumber \\
+\sum_{<i,j>^{\prime}}W_{i,j}(n_{i}-1)(n_{j}-1)-\mu\sum_{i}n_{i},
\end{eqnarray}

\vspace{0.1in}

where $n_{i\sigma}=c_{i\sigma}^{\dagger}c_{i\sigma}$ is a density operator
of the conduction electrons, $n_{i}=n_{i\uparrow}+n_{i\downarrow}$, $<ij>$
is a nearest neighbor pair and $<ij>^{\prime}$ is any other more distant pair. 
We measure energy in units of t and set t=1. 
For homogeneous states, the extended repulsive terms enters
only through W, which 
is the summation of $W_{ij}$ over all extended pairs $<ij>^{\prime}$
divided by the total number of lattice sites. For inhomogeneous state
calculations we need to specify the $W_{ij}$. To facilitate the
calculations, we consider only  next-nearest neighbor repulsive interactions. 
In Section IV the effect of a  dopant is represented by a short
range potential added to the Hamiltonian (1).

The tUVW model can be solved in the mean-field approximation by 
linearizing the interaction terms~\cite{mrr,mrr1}.
In linearizing the attractive V term,
one arrives at three linear terms. The first is  a density term, the second
is a pairing term and the third is an exchange term. As opposed to most
previous calculations, we retain all three terms. For the W term, only the
density term is kept after linearization. The linear Hamiltonian is
diagonalized by making a Bogoliubov transformation.
As usual, self-consistent solutions are obtained by
iterations. Spatially inhomogeneous solutions are readily obtained by
starting from inhomogeneous initial configurations.

\noindent {\bf III. Impurity Free Solutions}

All the calculations in this paper are done on a 24x24 lattice
at zero temperature. To see the
finite lattice effect and also to review the free energy argument for
phase separation, we show in Figure 1 the free energy as a function of
band filling on the finite lattice. The parameters used are the same as in I.
Phase separation should occur between
points A and B on the figure.
There is essentially no difference between the finite lattice
and infinite lattice results.

The parameters adopted in I (U=2.1, V=-0.94, W=0.4)
were chosen to reproduce the cuprate phase diagram. Empirically, we found
that the W term is essential 
for this purpose. As an illustration, we show in
Figure 2 the free energy curve for U=2.94, V=-0.9 and W=0. Phase separation
still occurs, with a terminal bandfilling 0.25,
but the d-wave superconductivity (DSC) is confined to inside the
two-phase region. It ceases to exist for band filling less than 0.275. In
contrast to the scenario presented in I, the phase separation boundary in this
tUV model does not cut through the superconducting dome. In other words,
homogeneous DSC does not exist in this case.

As an example of phase separation in the tUVW model, we choose an average
density (per spin) 0.4586
corresponding to 48 holes on the 24x24 lattice. In the
simple Hubbard model, two widely spaced stripes~\cite{zan,bish}
 (antiferromagnetic domain
walls) are expected. In contrast, the solution of the tUVW model as depicted
in Figure 3 is a narrow strip of low density (0.35) region embedded in a
half-filled region. The half-filled region has the maximum magnetization(0.48),
whereas the low density region has a nearly zero magnetization. The result thus
is consistent with a phase separation scenario.
By varying the band filling, we see similar results for band filling between
A and B in Figure 1, i.e. the free energy argument for phase separation
is explicitly demonstrated by  finite lattice calculations.

In this connection, we have also examined a bipolaron solution corresponding
to doping with two holes. As shown in Figure 4, there is a depression of the
SDW order parameter (and the electron density)
at the center of the bipolaron. This is expected because SDW is very stable 
at half-filling, any small change from half-filling is localized to a small
region. What is surprising is the large local DSC order paramter present
in the same neighborhood (Figure 5). This is a superconducting cloud surrounding a
SDW bipolaron. For technical reasons, we can not obtain a polaron solution.
We speculate that a polaron configuration would look qualitatively the
same as a bipolaron. Assuming that to be true, the above result
 implies that although
DSC is completely suppressed by SDW at half-filling in the ground state,
the DSC phase does show up in the charged excited states (the polarons
and bipolarons). This might explain the d-wave like dispersion of the
photoexcited states in the undoped cuprates~\cite{ronn}.

\noindent {\bf IV. Inhomogeneous Solutions in the Presence of Impurities}

As we have alluded to in the Introduction, in the two-phase region the
compressibility is zero because the free energy is a linear function of the 
density. That means an arbitrarily small perturbation can lead to significant
density variation. In particular, even a very weak dopant or impurity potential
can induce inhomogeneous solutions.

We model the impurity potential as follows: the potential at the impurity site
is take to be -0.35t, it reduces to -0.175t on the four nearest neighbors and
vanishes elsewhere. We constrain the separation between two impurities to be
larger than one lattice spacing. For a given distribution of impurities, the
sum of the potentials is added as a diagonal term to the Hamiltonian (1).
Only next-nearest neighbor pairs are considered for the repulsive W term.

For a typical case, we pick an average density of 0.417 per spin (96 holes).
Sixty impurities are randomly positioned on the lattice as shown in Figure 6.
Figure 7 depicts the variation of the DSC order parameter over the lattice.
The average value of the DSC order parameter is 0.049. Obviously, there
is a significant spatial variation (it varies from 0.01 to 0.073).
The SDW order parameter depicted in Figure 8 clearly exhibits a high plateau
region (half-filled region) and a low flat region (with zero magnetization). In
addition, there is a region of negative magnetization. A very weak impurity
potential (0.35t) seems to exert a strong influence in causing nanoscale
inhomogeneities. It should be emphasized again that according to Figure 1,
a homogeneous solution of this density (0.417) is superconducting but without
SDW. Instead, phase separation together with weak impurity potential lead to 
nanoscale inhomogeneous superconductor. This provides  the theoretical basis
of an interpretation of the STM and ARPES data~\cite{dell,chensu,fang,tam}
 of the cuprates.

\noindent {\bf IV. Discussion}

The main point of this paper is to demonstrate that inhomogeneous 
DSC and SDW are natural consequences of the interplay between them.
Both are derived from the UVW potential.
Impurities alone without phase separation
can also lead to inhomogeneous solutions as shown by other
 workers~\cite{wan,dag,atk,nunn}.
In our theory, the inhomogeneity is related to phase separation and
therefore occurs within the two-phase region (at least for
weak impurity potential). 

It is useful to look at the results of this paper from a broader perspective.
The attractive onsite Hubbard model has been employed as a standard model
for studying an s-wave superconductor. The model can explain most of the
superconducting properties while leaving the origin of the pairing force open.
That could be mediated by phonons, excitons or something else. In a similar vein,
the tUV model can be regarded as a standard model for a d-wave superconductor
independent of the origin of the attractive pairing force (the attractive V term).
Such a term could also generate s-wave superconductivity, therefore an onsite
repulsive U term is needed to suppress it. 

As is well known, the U term so introduced also leads to antiferromagnetism. The
V term also has important implications other than generating DSC, it can cause
phase separation~\cite{wppp,mrr,wang,mell}.
 The phase separation occurs at a temperature higher than the
superconducting transition temperature. In that sense, it is more important
than DSC. With the addition of an extended repulsive W term, 
the model can yield a phase diagram~\cite{wppp} strikingly similar to the one
observed in the cuprates~\cite{tim,tall}. 
In addition, inhomogeneous DSC~\cite{el} is readily explained.
From such a perspective, much of the mystery of the cuprate superconductors could
simply be a consequence of d-wave pairing in the appropriate parameter range,
again independent of the origin of pairing.

In an s-wave superconductor, one can separate those effects which are due
exclusively to electron-phonon coupling from those which are universally shared
between all s-wave superconductors. The same thing should be done for a d-wave
superconductor. It is pressing to figure out which portion of the cuprate
phenomenology is d-wave universal and which portion is specific to pairing
mechanism. That will help narrow the search for a pairing mechanism.
The specific form of the interaction potential UVW, with its alternate signs,
positive U and negative V plus positive W as the interparticle separation increases
strongly resembles the Friedel oscillation. As such it might hint at a 
particular pairing mechanism.

This work was partially supported by the Texas Center for Superconductivity, 
the Robert A. Welch Foundation (grant number E-1070),
and the National Science Council of Taiwan under contract number NSC
94-2811-M-110-005. I thank C. S. Ting for useful discussions. 
Thanks are also due to I. M. Jiang and the Department of Physics at the
National Sun Yat-sen University for their hospitality during the summer of
2005.

\begin{figure} 

\noindent Figure 1.
Free energy (per site)
of the homogeneous d-wave superconducting state (squares)
and that of the homogeneous coexisting SDW and DSC state (circles) at zero temperature in the
tUVW model.
\vspace{0.15in}

\noindent Figure 2. Same as Figure 1 for the tUV model.
\vspace{0.15in}

\noindent Figure 3. Charge density profile of a solution of the tUVW model
corresponding to an average density of 0.4586 per spin. The strip of low
density (0.35) region extends in the forward direction. The plane at the 
bottom is the lattice plane, the charge density corresponds to the height
of the surface above this plane.
\vspace{0.15in}

\noindent Figure 4. Spin-density-wave profile of a bipolaron.
\vspace{0.15in}

\noindent Figure 5. D-wave superconducting order parameter profile of a bipolaron.
The order parameter is defined as $\Delta_{d}=|\frac{V}{2}<c_{i\uparrow}c_{j\downarrow}
-c_{i\downarrow}c_{j\uparrow}>|$ 
(i,j is a nearest
neighbor pair)
The maximun of $\Delta$ at the center of the bipolaron is 0.064.
\vspace{0.15in}

\noindent Figure 6. Positions of sixty impurities randomly distributed on the lattice.
\vspace{0.15in}

\noindent Figure 7. Spatial variation of the DSC order parameter
of an inhomogeneous
solution with average density 0.416 per spin in the presence of sixty impurites
shown in Figure 6.
\vspace{0.15in}

\noindent Figure 8. Spatial variation of the SDW order parameter over the lattice
plane in the inhomogeneous solution described in Figure 7.
\vspace{0.15in}

%\begin{thebibliography}{999}
%\label{FIG:1}
\end{figure}

%\begin{figure} 

%\label{FIG:2}
%\end{figure}

\vspace{0.3in}

\end{document}